\documentclass[pre, twocolumn, amsmath, amssymb, superscriptaddress]{revtex4}
\usepackage{multirow}
\usepackage{graphicx}
\usepackage{dcolumn}
\usepackage{bm}
\usepackage{braket}
\usepackage{xcolor}

\usepackage{amsmath}

\newcommand{\beq}{\begin{equation}}
\newcommand{\eeq}{\end{equation}}

\usepackage{xr}
\usepackage{caption}

\usepackage{setspace}
\usepackage[a4paper,margin=25mm]{geometry}
\begin{document}

\title{Deciphering the chemical grammar of protein-RNA condensates}

\author{Greta Grassmann  \footnote{\label{corr} For correspondence write to: greta.grassmann@iit.it and mattia.miotto@uniroma1.it}}
\affiliation{Center for Life Nano \& Neuro Science, Italian Institute of Technology, Viale Regina Elena 261, Rome, Italy.}

\author{Giancarlo Ruocco}
\affiliation{Center for Life Nano \& Neuro Science, Italian Institute of Technology, Viale Regina Elena 261, Rome, Italy.}
\affiliation{Department of Physics, Sapienza University of Rome, Piazzale Aldo Moro, Rome, Italy.}

\author{Mattia Miotto \ref{corr}}
\affiliation{Department of Physics, Sapienza University of Rome, Piazzale Aldo Moro, Rome, Italy.}
\affiliation{Center for Life Nano \& Neuro Science, Italian Institute of Technology, Viale Regina Elena 261, Rome, Italy.}

\begin{abstract}
Biomolecular phase separation is typically attributed to the polymer physics of long, disordered chains. However, the underlying chemical grammar, i.e. the specific interactions between protein and RNA building blocks, remains poorly understood. We decouple those effects by screening the phase behavior of the complete dipeptide library in presence and absence of nucleic acids using full-atomistic molecular dynamics simulations. We demonstrate that (i) even these ultrashort units encode the instructions for spontaneous condensation, proving that phase separation is fundamentally rooted at a sub-polymeric level. (ii) Nucleic acids do not act as generic anionic glue but exert instead a base-specific regulatory logic. (iii) Individual nucleobases function as chemical tuners that dissolve, stabilize, or fluidize condensates based on their molecular identity. Overall, our minimal framework reveals that while polymer length enhances assembly, the core properties and regulatory control of condensates may be also governed by a fine-tuned chemical alphabet of peptides and nucleobases.
\end{abstract}

\maketitle

\section*{Introduction}
Compartmentalization of proteins and nucleic acids into membraneless assemblies known as biomolecular condensates has recently emerged as a key mechanism for cells organization~\cite{prouteau2018regulation, wagh2021phase} and function~\cite{banani2017biomolecular,shin2017liquid}, influencing a spectum of processes ranging from 
 signaling \cite{chattaraj2024multi} and regulation \cite{pei2025transcription, tian2020rna} to stress response~\cite{mccluggage2021paraspeckle, kuffner2021sequestration}  and noise buffering~\cite{simpson2021noise}. 
The formation of these condensates in living cells takes place by a phase separation process \cite{brangwynne2009germline}, in which a homogeneous solution demixes into two or more coexisting phases \cite{hyman2014liquid}. 
Deficiency of the cells in regulating such demixes, i.e. in forming and dissolving condensates, typically translates in severe pathological conditions~\cite{iadanza2018new,li2022mechanism,shen2020biomolecular,ranganathan2022physics,szala2023challenges,grassmann2021computational}.
Thus, an ever increasing effort is being dedicsted to determine what  the physico-chemical ingredients regulating condensates dynamics are.
In this context, several studies have highlighted how cross-$\beta$ (amyloid-like) interactions can underlie protein self-assembly and, in some cases, contribute to condensate maturation \cite{alberti2021biomolecular,hughes2018atomic,kato2012cell,luo2018atomic}. More generally, biomolecules capable of engaging in noncovalent intermolecular interactions (e.g. electrostatic, cation–$\pi$, $\pi$–$\pi$, and hydrophobic interactions) can drive this process. Intrinsically disordered proteins (IDPs) are often key drivers of phase separation due to their conformational heterogeneity and multivalent interaction capacity \cite{brangwynne2015polymer,van2014classification,wright2015intrinsically}.
From a chemical view point, sequences promoting demixing are frequently enriched in polar, charged, or aromatic residues
and often contain short low-complexity motifs such as YG/S-, FG-, RG-, GY-, KSPEA-, SY-, or Q/N-rich regions, as well as blocks of alternating charge \cite{brangwynne2015polymer},
suggesting that even minimal interaction motifs may encode essential physicochemical principles of assembly formation \cite{nguyen2022condensates,grassmann2023electrostatic}.
RNA molecules found within protein condensates \cite{roden2021rna} are known to  promote or hinder phase separation of condensates via two major ingredients, i.e. their net negative charge and  structural flexibility   \cite{pak2016sequence,langdon2018mrna,ries2019m6a,elbaum2015disordered,zhang2015rna,maharana2018rna}. 
Notably, short nucleic acid fragments, including aptamers, have also been reported to modulate or disrupt condensation \cite{huang2024inhibition,matos2020liquid}, 
 suggesting that the two features can be uncoupled.\\

A wide range of coarse-grained (CG) molecular dynamics (MD) models have been developed for proteins and RNA to investigate their essential structural properties (such as flexibility) and their effect on phase separation \cite{liu2025toward,lebold2022tuning,yasuda2025coarse}.
To elucidate the chemical aspect on the other hand, minimal molecular systems retaining essential interaction motifs are being investigated. Recent studies have shown that low-molecular-weight compounds, including short peptides and small molecules, can self-associate and even form ordered assemblies \cite{bao2025phase,abbas2021short}.
Among peptides, dipeptides represent the minimal sequence length capable of experimentally observable self-association \cite{frederix2011virtual}.\\
Just as amino acid composition influences protein phase behavior, the nucleotide composition of RNA affects its condensation properties. Purines -adenine (A) and guanine (G)- and pyrimidines -cytosine (C) and uracil (U)- confer distinct interaction capabilities \cite{roden2021rna, bang1910untersuchungen}. Subsequent experimental studies confirmed that RNA homopolymers can phase separate and that nucleotide composition modulates condensate properties \cite{li2012phase}.\\

In this study, we first employed all-atom molecular dynamics (MD) simulations to systematically screen all possible dipeptide combinations encoded by the twenty standard amino acids and to predict their propensity to undergo phase separation. 
Next, we moved to investigate how this behavior is modulated by the presence of nitrogenous bases, mimicking short RNA fragments found in cellular condensates.
By comparing the behavior of each dipeptide in the absence and presence of individual nitrogenous bases, we aim to determine whether nucleobases generally enhance, suppress, or qualitatively transform peptide-driven condensation, and whether these effects depend on residue chemistry.\\
To characterize phase separation propensity, we developed an efficient two-step framework: (1) condensation capability is quantified through clustering degree and collapse degree of initially dispersed dipeptides; and (2) structural and dynamical descriptors are used to characterize the resulting clusters. Even this minimal model recapitulates established principles of protein phase separation observed in larger biomolecular condensates, including the dominant role of hydrophobic and $\pi$-mediated interactions.\\
Applying the same protocol in the presence of nucleobases reveals a threefold behavior: nitrogenous bases can dissolve dense clusters, promote more dynamic and weakly associated assemblies, or preserve highly packed clusters while altering their dynamics and persistence. These findings parallel proposed cellular mechanisms by which RNA tunes the material properties of protein condensates \cite{maharana2018rna}. Moreover, we observe a pronounced dependence on nucleobase identity, highlighting that RNA’s role in condensates extends beyond generic multivalency or sequestration effects. This base-specific modulation may have implications for the rational design of RNA- or small-molecule-based strategies aimed at controlling aberrant protein aggregation.

\section{Results}

\begin{figure*}[]
\centering
\includegraphics[width=\linewidth]{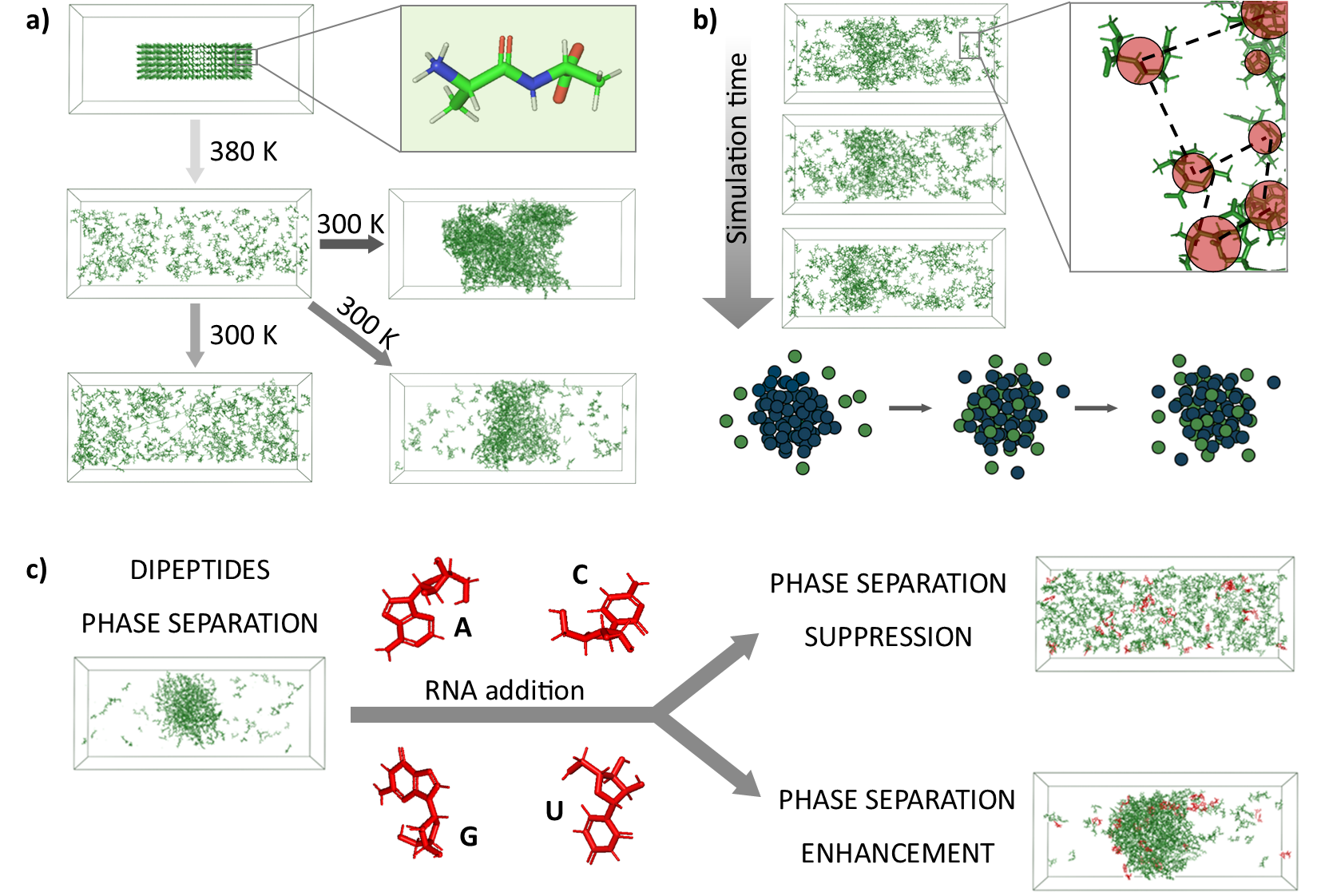}
\caption{\textbf{Minimal model for protein-RNA condensate formation.} 
Summary of the simulation pipeline. \textbf{a)} Systems are initialized from crystallographic-like structures. An initial all-atom simulation at 380 $K$ perturbs the system, followed by a 300 $K$ simulation to evaluate condensate formation or dispersion. 
\textbf{b)} Static and dynamic descriptors are computed from the trajectories of each dipeptide’s center of mass, and clusters are identified based on inter-dipeptide distances.
\textbf{c)} Simulations in presence and absence of the four nitrogenous bases are performed to investigate their influence on dipeptide phase separation.}
\label{figintro}
\end{figure*}

To isolate the effect of chemical composition in protein-RNA condensates, we performed an extensive campaign of all-atom standard molecular dynamics simulations with explicit solvent probing all combinations of protein dipeptides and RNA nucleobases mixtures for a total of 2000 different systems. In particular, we carried on a first set of simulations at high dipeptide concentration ($\sim$0.9M) and a null concentration of nucleobases to determine the aggregation propensities of the amino acids couples in absence of RNA. Then, we carried out on a second set of four simulations for each dipeptide combination, each including one of the four nitrogenous bases species with a peptide-nucleobasis concentration of 10:1. 
For each system, we first run a $\sim$50 $ns$-long simulations at 380 K to remove possible biases due to the disposition of the moleculars in the initial configuration, followed by  a $\sim$50 $ns$-long simulations at 300 K (see Figure~\ref{figintro}a and Methods for details).
The trajectories of the dipeptides’ centers of mass were tracked and analyzed in terms of clusters formation and compositional dynamics (Figure \ref{figintro}b) to distinguish dipeptides with phase separation propensity and characterize their different behaviors in 
the absence or presence of the nitrogenous bases (see Figure \ref{figintro}c).

\subsection*{Determination of the phase separation propensity of dipeptide condensates}
\begin{figure*}[]
\centering
\includegraphics[width=\linewidth]{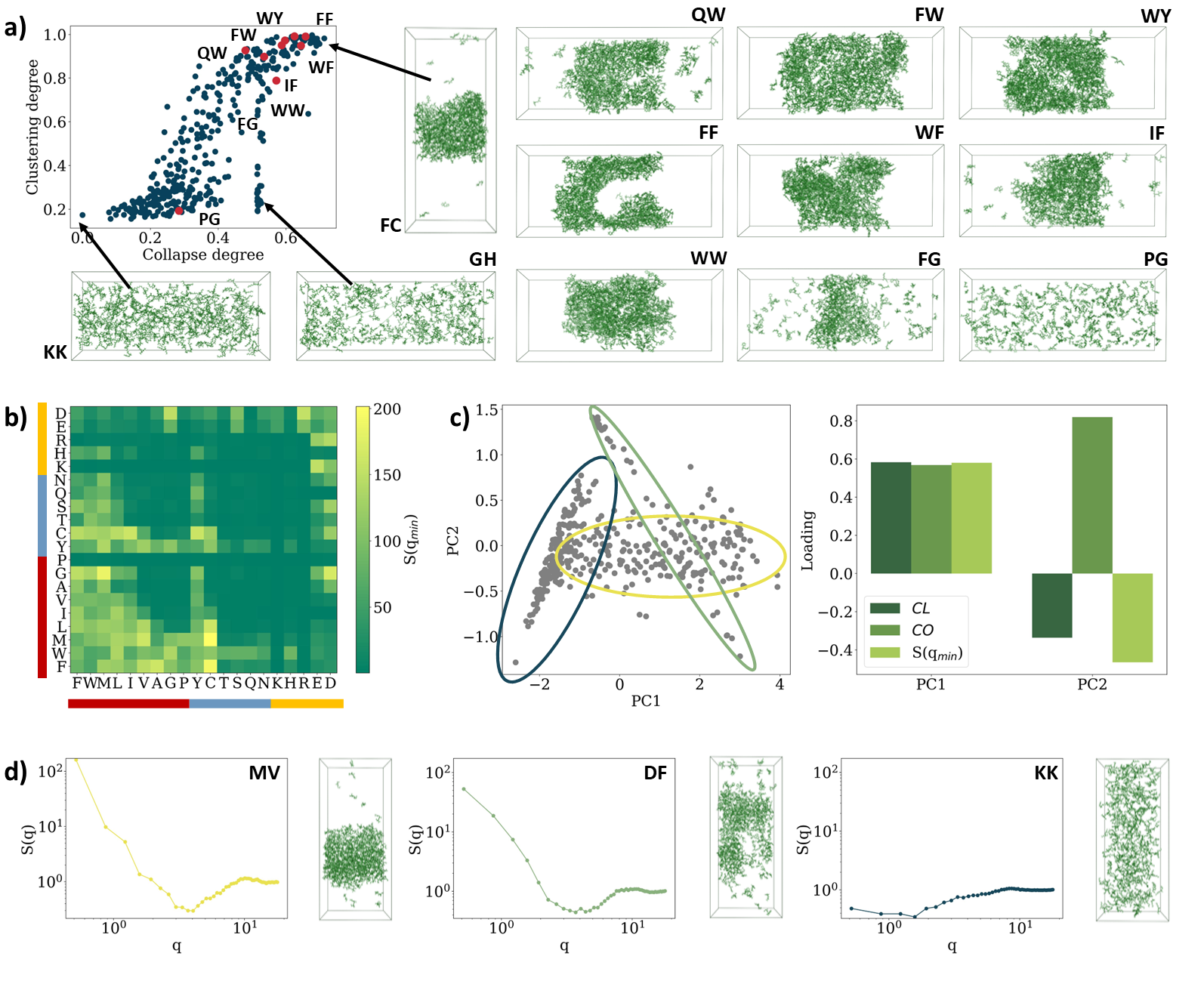}
\caption{\textbf{Identification of dipeptide phase separation tendency.} 
\textbf{a)} Clustering degree versus Collapse degree for each dipeptide combination. Red points indicate dipeptides with experimental evidence of phase separation or aggregation. Surrounding boxes show the last simulation frame of dipeptides with the highest (KK) and lowest (FC) values, and an example with average behavior (GH). On the right, equilibrium snapshots for the experimentally characterized dipeptides are shown. 
\textbf{b)} Heatmap of the $S(q_{min})$ value computed for each dipeptide combination (as indicated by the colorbar). Residues are grouped by chemical nature (hydrophobic, polar, and charged residues in red, light blue, and yellow, respectively).
\textbf{c)} For each simulation, Clustering degree, Collapse degree, and $S(q_{min})$ are computed. A PCA is performed on the dataset, which is then projected on the first two PCs. Three ellipses are drawn in correspondence of points following similar trend. The yellow ellipse highlight the regions of dipeptides that tend to form solid and dense clusters. The blue ellipse corresponds to dipeptides that stay dispersed in the solution. The green ellipse correspond to dipeptides that tend to form intermediate or more dispersed assemblies. On the right, the loadings of each PC.
\textbf{d)} Representative simulation final frames and $S(q)$ (MV, DF, KK) illustrate how each class correspond to changes in compactness and spatial organization.}
\label{fig1}
\end{figure*}

To begin with, we focused on the 400 dipeptides combinations in absence of RNA and quantified the aggregation propensity of the systems. 
For each frame of the trajectory, dipeptides whose centers of mass were separated by less than $C_{t} = 10$ \AA were assigned to the same cluster. The threshold was chosen consistent with typical first-shell interaction distances in short peptide clusters \cite{hansen2013theory}.
To quantify phase behavior, we defined two global descriptors: the Clustering degree ($CL$) and the Collapse degree ($CO$). The Clustering degree is computed as the average number of dipeptides in the largest cluster in each frame over the total number of dipeptides: high $CL$ values indicate collective self-assembly involving a substantial fraction of molecules. The Collapse degree is defined as $CO = 1 - SASA_f$, where $SASA_f$ is the Solvent-Accessible Surface Area (SASA) averaged over the last ten frames and normalized by the maximum SASA observed across all simulations. Because SASA inversely correlates with molecular burial and intermolecular packing, high $CO$ values reflect sequestration from solvent and dense packing.
Figure \ref{fig1}a shows the relationship between $CO$ and $CL$ for all 400 systems. A strong positive correlation emerges (Pearson $r = 0.82$, $p < 10^{-98}$), indicating that molecular compaction and collective clustering are tightly coupled. Systems that collapse extensively also tend to recruit a large fraction of molecules into dominant assemblies.
Interestingly, dipeptides populate the $CL$–$CO$ plane in a sigmoidal distribution, with over representation at the extremes: either highly collapsed, densely clustered systems or dispersed ones. Intermediate states are comparatively less frequent, suggesting a cooperative transition between dispersed and clustered regimes, reminiscent of nucleation-like behavior observed in phase separating systems \cite{sear2007nucleation}.

To test our pipeline, we compared our predictions with available experimental data. In particular, we retrieved all known dipeptides that have been observed to form condensates (see Table \ref{tab1}) from literature. 

\begin{table}[]
\centering
\begin{tabular}{|c|c|}
\hline
Dipeptide & Reference(s) \\
\hline
FF & \cite{reches2003casting,yan2010self}  \\ 
FG & \cite{kumar2016aggregation}  \\
FW & \cite{reches2003casting} \\
IF & \cite{reches2003casting,de2007ile}  \\
PG &  \cite{kumar2016aggregation} \\
WF & \cite{reches2003casting} \\
WY & \cite{reches2003casting} \\
WW & \cite{reches2003casting} \\
QW & \cite{tang2021prediction} \\
\hline
\end{tabular}
\caption{\textbf{Experimentally validated phase-separating dipeptides.} List of dipeptide systems reported to undergo condensate formation, along with the corresponding references.}
\label{tab1}
\end{table}

All but one of the dipeptides with experimental evidence of phase separation or aggregation localize in the high-$CL$, high-$CO$ region (see red dots in Figure~\ref{fig1}a).
Notably, such results confirm that relatively short atomistic simulations can be used to capture intrinsic phase separation propensity.
Only one dipeptide experimentally known to cluster occupies the low-$CL$/low-$CO$ region, PG \cite{kumar2016aggregation}, which in our simulations does not form detectable clusters. 
We hypothesize that this discrepancy may arise from the conformational restriction imposed by proline (P) \cite{ganguly2020conformational}, which limits backbone flexibility, removes an amide hydrogen donor, and thus constrains intermolecular packing in our simulation condition.

To quantify mesoscale organization, we computed the static structure factor $S(q)$ \cite{hayter1981analytic} for each simulation (see Methods for more details). In the long-wavelength limit, the structure factor obeys the compressibility relation: $S(q\rightarrow0)=\rho k_B T \kappa_T$, where $\rho$ is the density, $k_B$ the Boltzmann constant, $T$ the temperature, and $\kappa_T$ the isothermal comprimibility \cite{hansen2013theory}. Thus, $S(q \rightarrow 0)$ directly probes large-scale density fluctuations and provides a signature of collective organization: its increase reflects enhanced compressibility and mesoscale density heterogeneity, hallmark features of phase separating systems \cite{cerda2004structure, arxiv.2411.08012}.
Figure \ref{fig1}b presents the residue-resolved heatmap of $S(q_\text{min})$. Aromatic and hydrophobic residues dominate the high-density region, consistent with the central role of hydrophobic and $\pi$-stacking interactions in promoting interactions \cite{martin2020valence, vernon2018pi,desantis2022spatial}.
Charged and polar residues contribute lower values, suggesting reduced spontaneous phase separation propensity under the simulated conditions.\\

To collect the essential information brought by these descriptors, we performed a Principal Component Analysis (PCA) using $CL$, $CO$, and $S(q_{min})$ as input variables, and projected the dataset on the first two PCs (see Figure \ref{fig1}c), explaining 95\% of the total variance.
Three broad classes are revealed, distinguished especially on the first PC, where the three variables have similar loadings.
The three groups tend to correspond to i) highly collapsed, dense assemblies, ii) intermediate, dynamically reorganizing clusters, and iii) dispersed systems lacking persistent organization.
Systems in class i) exhibit markedly higher $S(q_\text{min})$ values, reflecting enhanced density fluctuations and reduced compressibility associated with condensed phases. In contrast, class iii) systems show minimal long-wavelength density correlations. Figure \ref{fig1}c report an example of the $S(q)$ and final simulation frame for one dipeptide from each group.\\

These results demonstrate that our minimal model captures the basic principles of protein phase separation. We observed that dipeptide phase behavior is strongly sequence-encoded: local chemical interactions modulate packing density, which in turn determines long-wavelength density fluctuations.

\subsection*{Sequence-dependent packing determines cluster morphology and dynamics}

\begin{figure*}[]
\centering
\includegraphics[width=\linewidth]{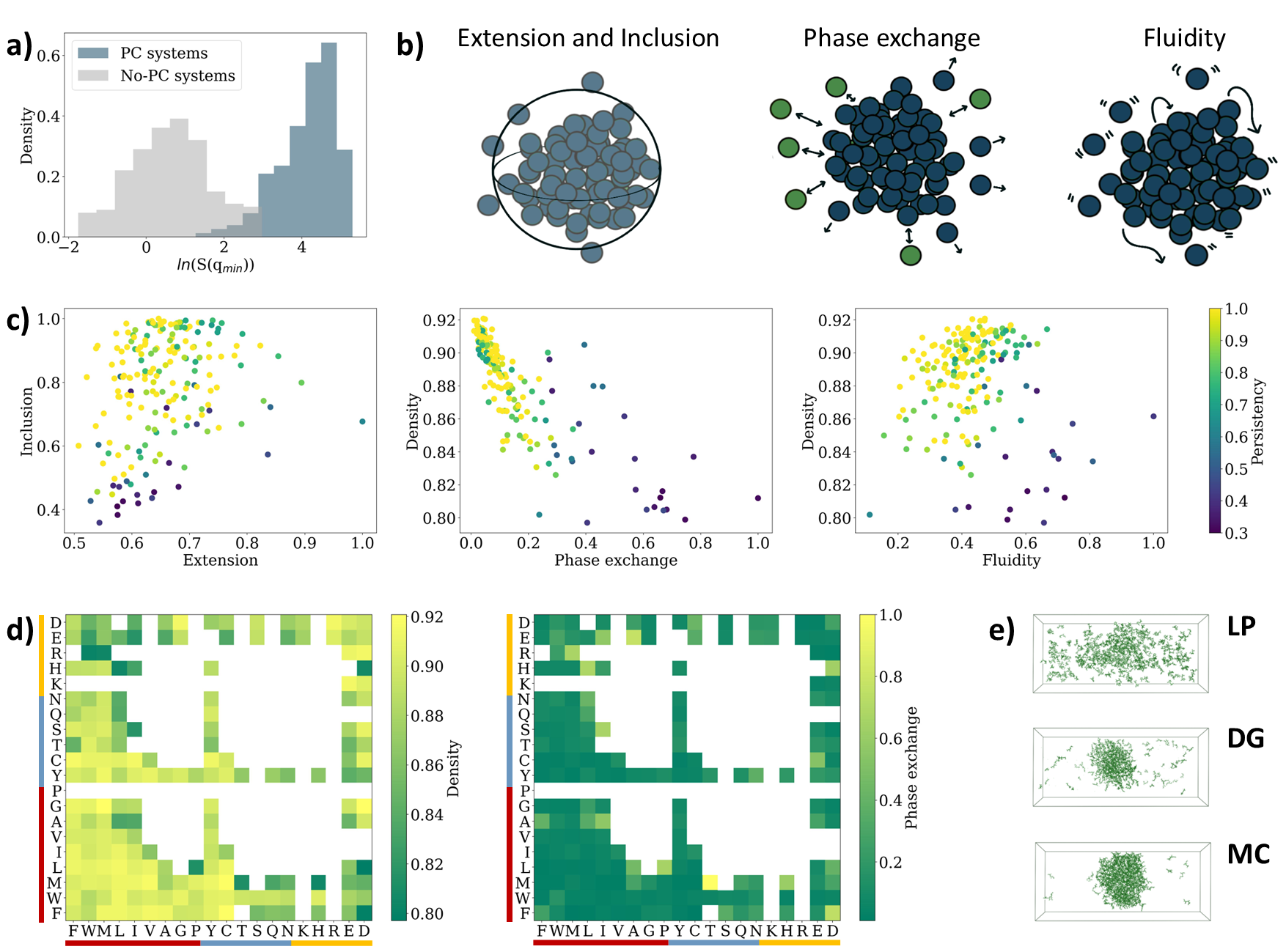}
\caption{\textbf{Structural and dynamical characterization of Persistent Clusters (PCs).} 
\textbf{a)} Distribution of the logarithm of the $S(q_{min})$ values for $PC$-forming and non-$PC$ systems.
\textbf{b)} Graphical representation of the definition of Extension, Inclusion, Phase Exchange, and Fluidity.
\textbf{c)} On the left, Inclusion versus Extension for all $PCs$. In the middle, Density versus Phase Exchange, on the right Density versus Fluidity.
 Points are colored according to the Persistency of the $PC$.
\textbf{d)} Residue-resolved maps of Density (left) and Phase Exchange (right), with residues grouped by chemical nature (hydrophobic, polar, and charged residues in red, light blue, and yellow, respectively). 
\textbf{e)} Representative final configurations are displayed for residue pairings that deviate from the typical behavior of their respective chemical classes in d.}
\label{fig2}
\end{figure*}

To analyze the systems exhibiting phase separating behavior within our model, we searched for the largest cluster in each simulation persisting for more than 30\% of the total simulation time. We refer to such assemblies as Persistent Clusters ($PCs$). This operational definition selects agglomerates that are both spatially extensive (the smallest $PC$ having on average more than 35\% of the system dipeptides) and temporally stable, consistent with the physical criteria typically used to define condensed phases \cite{hyman2014liquid}.
Systems that do not form $PCs$ exhibit low values of the static structure factor in the long-wavelength limit $S(q_{min})$ (mean = 3.19), whereas $PC$-forming systems display markedly higher values (mean = 77.39). Indeed, such separation between $PC$-forming and non-forming systems results in two non-overlapping distributions of $S(q_{\text{min}})$  (ROC AUC = 0.99), as shown in Figure \ref{fig2}a. 
Accordingly, systems lacking $PCs$ were classified as non–phase separation-prone.\\
For each $PC$, we quantified three structural descriptors: Inclusion ($I$), Extension ($E$), and Persistency ($P$).
Inclusion is defined as the maximum-normalized average number of dipeptides in the cluster, while Extension corresponds to the maximum-normalized average radius of gyration ($R_g$) (Figure \ref{fig2}b and the Methods Section). Persistency represents the normalized lifetime of the $PC$.
Figure \ref{fig2}c shows that short- and long-lived $PCs$ tend to span similar spatial extension, but the former incorporate fewer molecules. Stable $PCs$ achieve both higher molecular inclusion and sustained temporal stability, suggesting cooperative stabilization mechanisms.\\

To characterize the material properties of the condensates, we analyzed both internal mobility and exchange with the surrounding environment. These analyses were complemented by a density-like descriptor defined as $D=1-\frac{R_g}{size}$. Although $D$ does not represent a thermodynamic density, it provides a comparative measure of spatial compactness across clusters of different sizes, with higher values indicating more tightly packed assemblies. As shown in Figure \ref{fig2}c, $PCs$ with greater Persistency tend to exhibit higher $D$, consistent with theoretical models and experimental observations linking interaction strength and network connectivity to condensate density and stability \cite{banani2017biomolecular,dignon2018sequence}.\\
Inspired by Fluorescence Recovery After Photobleaching (FRAP), a standard technique for probing molecular mobility in living cells \cite{meyvis1999fluorescence}, we defined a Phase Exchange ($PE$) parameter. 
$PE$ is calculated as the number of dipeptides that exit and re-enter the cluster at least once, divided by the average cluster size and lifetime, and then max-normalized. Higher 
$PE$ values indicate extensive molecular exchange and, therefore, more dynamic, liquid-like behavior. The center panel of Figure \ref{fig2}c shows that $D$ correlates negatively with Phase Exchange (Pearson $r=-0.81, p<10^{-45}$); denser clusters exchange fewer molecules with the surrounding solvent, whereas less dense clusters exhibit higher exchange, contributing to their erosion and reduced Persistency.\\
To probe internal dynamics, we defined Fluidity ($FL$) as the maximum-averaged root mean square deviation (RMSD) of dipeptides that remain within a cluster for its entire lifetime, normalized by its lifetime. As observed for $PE$, less dense and short-lived clusters tend to be more internally dynamic. However, no significant correlation was observed between Density and Fluidity, suggesting that internal mobility alone is not sufficient to predict cluster stability.\\
Figure \ref{fig2}d presents residue-resolved maps of $PCs$ Density and Phase Exchange.
The maps reveal sequence-dependent trends, indicating that the phase behavior of dipeptides is strongly encoded by amino-acid chemistry, and the density map mirrors the trends observed for $S(q_{min})$.\\
Hydrophobic–hydrophobic combinations (H-H) tend to display the highest Density and the lowest Phase Exchange, highlighting the dominant role of hydrophobic interactions in promoting compact cluster organization. There are some exception: LP forms a particularly less stable $PC$ with more exchange with the solvent, potentially because of the structural rigidity of proline and/or the lack of a backbone NH group, limiting hydrogen-bonding capability. A similar behavior is also shown by other hydrophobic pairs, such as AL, AI, and LA. In these cases, differences in size and branching (e.g., small Ala vs bulky Leu/Ile) could lead to inefficient packing and local free volume within the cluster, reducing overall density. Hydrophobicity is not sufficient to ensure high compactness and low exchange: efficient packing, side-chain geometry, and backbone conformational constraints critically modulate cluster stability.
As observed in literature, aromatic residues (F, W, Y) often display high Density and low Phase Exchange, consistently with the stabilizing role of $\pi$–$\pi$ stacking and cation–$\pi$ interactions \cite{brangwynne2015polymer}, which enhance intermolecular cohesion and favor the formation of dense phases. This behavior changes when they are combined with charged residues (C).
Charged residues generally correspond to lower packing density and also enhanced molecular exchange. Electrostatic repulsion disfavors compact organization unless compensated by complementary interactions. Indeed, an example of a more stable $PCs$ is DR (negative-positive charge).
Polar residues (P) can promote lower densities and higher exchange, indicating weaker cooperative stabilization within the condensed phase. Some exception are the system including a polar and a hydrophobic residue, such as MC and FC. In these cases, the hydrophobic residue could nucleate a compact core through dispersion-driven interactions, while the polar residue either remains interfacially exposed or contributes secondary interactions without significantly increasing free volume.\\

\subsection*{Nitrogenous bases rewire dipeptide condensate by modulating phase separation and condensate persistence in a threefold fashion}

Next, we analyzed simulations in presence of each canonical nitrogenous base.
Figure \ref{fig3}a illustrates, as a representative example, the effect of uracil addition (see Supplementary Information for all nitrogenous bases). The presence of nitrogenous bases can exert a dual role on peptide condensation, leading either to an enhancement or to a suppression of phase separation.
 While some systems show a minimal effect, a substantial fraction exhibits significant deviations in $CL$ or $CO$.\\
Stratification by residue chemistry, averaged on all nitrogenous bases (Figure \ref{fig3}b), reveals that the effect is strongly sequence-dependent. Hydrophobic–hydrophobic combinations tend to show modest shifts, whereas the phase separation of systems including charged residues tends to be strongly reduced by nitrogenous bases. Systems including polar residues in some cases show instead an enhancement of phase separation propensity.
Such dual behavior mirrors the enhancement or dissolution of protein condensates by RNA reported in ribonucleoprotein systems \cite{elbaum2015disordered,zhang2015rna,maharana2018rna}.
The same analyses performed for the single nitrogenous bases can be found in the Supplementary.\\
Residue-resolved maps of the variation of $S(q_{min})$ following the addition of nitrogenous bases (Figure \ref{fig3}c) further demonstrate that they modulate long-wavelength density fluctuations in a chemically selective manner. Hydrophobic dipeptides frequently display negative variation, indicating enhanced large-scale density heterogeneity and increased phase separation propensity. In contrast, polar and charged combinations exhibit positive shifts. 
Notably, the magnitude of the effect can depend on the identity of the nucleobase, reflecting differences in hydrogen-bonding patterns, stacking strength, and electronic distribution among adenine, cytosine, guanine, and uracil. 
In addition, variations in steric bulk and molecular geometry lead to distinct excluded-volume contributions, which can further modulate the phase separation propensity of the system \cite{grassmann2024computational}.
For instance, the phase separation propensity of the CY system is strongly suppressed upon addition of adenine, whereas the same system supplemented with the other three nucleobases still forms a dense and extended $PC$.\\
Despite noticeable changes in the overall density distributions upon nucleobase addition, the internal density of $PCs$ that remain stable is only moderately affected. This behavior is exemplified by the case of uracil in Figure \ref{fig3}d1, while the corresponding analyses for the other nitrogenous bases are reported in the Supplementary Information.
Persistency is more strongly modulated than Density, with some $PCs$ displaying enhanced stability while others show increased turnover. Figure \ref{fig3}d2 illustrates this behavior for uracil as a representative example, whereas the effects of the other nitrogenous bases are reported in the Supplementary Information.
The distinct behavior of Density and Persistency points to a decoupling between structural compactness and dynamical stability, indicating that nucleobase addition reshapes the lifetime and exchange properties of $PCs$ more strongly than their internal packing. This aligns with observations that once a condensed phase is stabilized by a sufficiently connected interaction network, its local structure and density can be resilient to moderate changes in component chemistry \cite{espinosa2020liquid}.
Nevertheless, nucleobase addition substantially reshapes the landscape of possible clusters by selectively promoting or destabilizing specific assemblies. $PCs$ that are dispersed by nucleobase addition often exhibit intermediate Density prior to disruption and often have high Persistency, indicating that nucleobases do not simply perturb weakly packed systems but instead alter the interaction network in a sequence-specific manner, competing with existing peptide–peptide contacts \cite{krainer2021reentrant}. 
Conversely, in some systems that do not spontaneously phase separate in the absence of nucleobases, the addition of nitrogenous bases can promote lower-Density $PCs$ with both low or high Persistency. This observation supports a growing body of work showing that RNA and RNA-like molecules can act as condensate modulators, either enhancing or suppressing demixing depending on sequence and stoichiometry, by providing additional interaction sites or modifying the effective interaction connectivity of the system \cite{elbaum2015disordered,zhang2015rna,maharana2018rna,huang2024inhibition,matos2020liquid}.
All these mechanism could underlie the action of RNA as a tunable regulator of condensate material properties, which can have important role in physiological and pathological processes \cite{maharana2018rna}.\\
These effects are strongly sequence-dependent and vary across amino acid classes. Figures \ref{fig3}e1–e2 present the case of uracil as a representative example, while the corresponding results for the other nitrogenous bases are provided in the Supplementary.
Systems containing charged residues are the most perturbed by nucleobase addition, consistent with the charged character of the bases, and in many cases completely lose their ability to form $PCs$. Hydrophobic pairs that phase separate in the absence of nucleobases generally retain this ability after base addition, although they often exhibit reduced Density and Persistency. Interestingly, in a subset of systems, nucleobase addition enhances phase separation, indicating favorable heterotypic interactions. In polar–polar systems, these opposing tendencies appear with the same frequency.
The same analyses performed for individual nitrogenous bases are reported in the Supplementary.\\
Overall, these results demonstrate that nitrogenous bases act as chemically specific modulators of peptide-driven phase separation. Rather than exerting a uniform crowding effect, they selectively rewire interaction networks in a sequence-dependent manner, enhancing condensation in some dipeptides while destabilizing others. This behavior parallels the role of RNA and small metabolites in cellular condensates.

\begin{figure*}[]
\centering
\includegraphics[width=\linewidth]{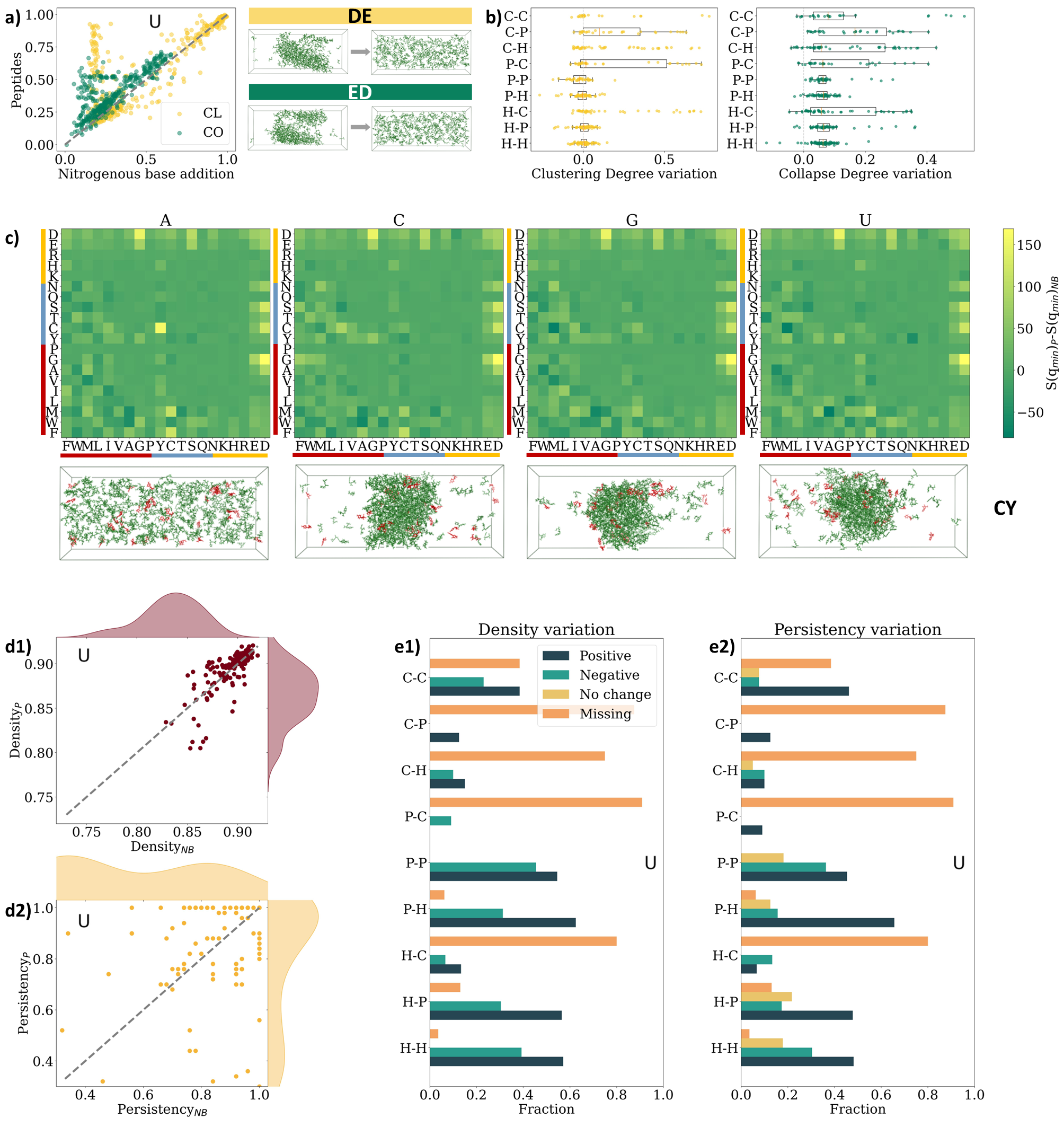}
\caption{\textbf{Nucleobase-specific modulation of dipeptide clustering, density fluctuations, and condensate stability.} 
\textbf{a)} Comparison of Clustering Degree and Collapse Degree in the presence (x-axis) and absence (y-axis) of uracil.
The dashed line indicates equality. 
Representative snapshots illustrate systems where base addition suppressed the Clustering Degree (DE) or the Collapse Degree (ED).
\textbf{b)} Distribution of changes in Clustering Degree and Collapse Degree upon nitrogenous bases addition, stratified by residue classes ($C$ = charged, $P$ = polar, $H$ = hydrophobic). Positive values indicate enhanced association in the presence of nucleobases. The values for the nitrogenous bases are averaged over the four cases for each dipeptide.
\textbf{c)} Residue-resolved heatmaps of the variation in long-wavelength density fluctuations for adenine (A), cytosine (C), guanine (G), and uracil (U). Positive values indicate reduced phase separation propensity upon base addition.
Under each heatmap the final frame of the CY system in each condition is reported.
\textbf{d1)} Comparison of Density in the presence (x-axis) and absence (y-axis) of uracil. Only the systems forming Persistent Cluster in both conditions are compared. The dashed line indicates equality. The marginal distribution on the top (right) axis indicate the Density values of the $PCs$ forming only in the systems with (without) uracil.
\textbf{d2)} Same as in d1, but for Persistency.
\textbf{e1)} Categorical variation of Density across residue classes in the presence of uracil. Dipeptides system that completely lose the propensity to phase separate following nitrogenous bases addition are indicated by the orange bar ("Missing").
\textbf{e2)} Same as in e1, but for Persistency.
}
\label{fig3}
\end{figure*}

\subsection*{Nitrogenous bases show different influence on cluster dynamics depending on their composition and structure}
Having established how nitrogenous bases modulate cluster packing and persistence, we next investigated their impact on cluster dynamics.\\
As shown in Figure \ref{fig4}a1 for the specific case of uracil (see Supplementary Information for the other bases), nucleobase addition disrupts several $PCs$ characterized by low Phase Exchange, confirming that nitrogenous bases can destabilize tightly packed clusters formed in their absence. This behavior is consistent with previous reports showing that RNA components can fluidize or remodel peptide condensates \cite{conti2022biomolecular}. 
In parallel, nucleobase addition can also promote $PCs$ with medium or high Phase Exchange, confirming that peptide–RNA interactions can also enhance molecular turnover within condensates.
A subset of $PCs$ retains low Phase Exchange under both conditions, but in some cases the addition of nitrogenous bases can have a detrimental or enhancing effect to the exchange with the environment.
To quantify variations in the internal mobility of $PCs$, we computed the Fluidity of the $PCs$ formed in the presence of nitrogenous bases.\\
Figure \ref{fig4}a2 shows that $PC$ Fluidity is markedly altered for systems that form $PCs$ under both conditions, with no clear correlation between the two. The addition of nitrogenous bases disrupts phase separation irrespective of whether the original $PCs$ exhibit high or low fluidity. At the same time, nucleobase addition promotes the formation of $PCs$ characterized by high internal mobility.
These dual behaviors align with the emerging view that RNA can act both as a scaffold promoting phase separation and as a solubilizing agent that buffers excessive aggregation, depending on concentration and sequence context \cite{pak2016sequence,langdon2018mrna,ries2019m6a,elbaum2015disordered,zhang2015rna,maharana2018rna,huang2024inhibition,matos2020liquid}.
Figure \ref{fig4}b summarizes the features of $PCs$ that are conserved (top), disrupted (center) or created (bottom) upon nitrogenous bases addition.\\
Figures \ref{fig4}c-d reveal that these effects are strongly nucleobase-dependent. For example, adenine disrupts AM $PC$ formation while preserving RW and RM interactions and promoting $PC$ formation in NQ pairs. Cytosine hinders $PC$ formation in several polar–hydrophobic pairs but promotes it in GC. Guanine broadly increases Phase Exchange, even in hydrophobic–hydrophobic systems, yet preserves MF interactions, which are instead lost upon cytosine addition. Uracil promotes $PC$ formation in polar–polar pairs, in agreement with its comparatively weaker stacking propensity and greater reliance on hydrogen-bond-mediated interactions \cite{cyranski2003aromatic}.\\
Figure \ref{fig4}e summarizes the effect of the different nitrogenous bases, stratified according to the dipeptides classes. 
A, C, and U show trends similar to peptide-only systems,
where hydrophobic–hydrophobic pairs preferentially form $PCs$ with low Phase Exchange. This suggests that in these cases peptide–peptide hydrophobic contacts remain dominant over peptide–RNA interactions.
In contrast, G softens this trend and generally enhances Phase Exchange, including for H–H pairs, but the most notable case being the C-H pairs. This behavior may reflect guanine’s strong aromaticity and stacking propensity \cite{cyranski2003aromatic}. 
An important exception is represented by P–C pairs, which display higher Phase Exchange in the presence of U or C, indicating that subtle differences in hydrogen-bonding capability and electrostatic patterning among nucleobases significantly reshape interaction networks.
Despite exhibiting lower Phase Exchange, H–H pairs tend to display comparable or even higher Fluidity values than other classes, particularly in the presence of adenine. Adenine also enhances the internal mobility of H–P and C–H pairs more effectively than the other bases. Notably, uracil enhances phase separation in P–C pairs, leading to the formation of new $PCs$.\\
Overall, these results reinforce the concept that both dipeptide composition and RNA base identity critically determine phase exchange behavior: subtle chemical differences at the nucleobase level can tune the balance between stabilization and fluidization in peptide-driven phase separation.

\begin{figure*}[]
\centering
\includegraphics[width=\linewidth]{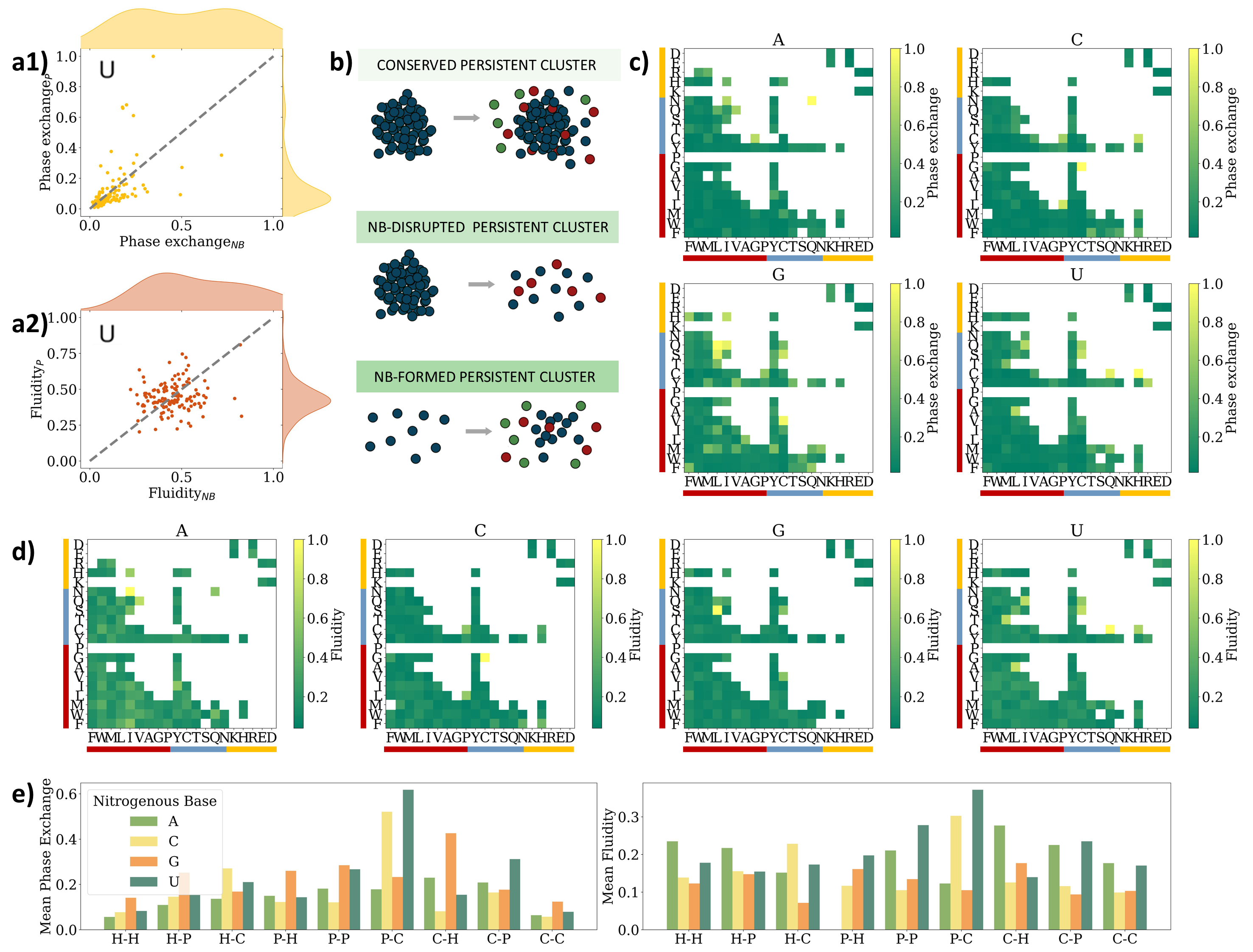}
\caption{\textbf{Nucleobase-dependent modulation of condensate exchange and internal dynamic.} 
\textbf{a1)} Scatter plot comparing the Phase Exchange of $PCs$ formed by peptides alone and in the presence of nitrogenous base U addition. The dashed line indicates parity. Only the system forming $PCs$ in both conditions are considered. The marginal histogram reports the distributions of the Phase Exchange values of the $PCs$ lost with the addition of nitrogenous bases.
\textbf{a2)} Same as in a1 for Fluidity.
\textbf{b)} Features of Persistent Clusters conserved (top), disrupted (middle) or formed (bottom) by nitrogenous bases addition.
\textbf{c)} Phase Exchange residue-resolved matrices for systems containing A, C, G, or U (from left to right, top to bottom). Residues are grouped in hydrophobics (red), polar (blue), and charged (yellow).
\textbf{d)} Same as in c but for Fluidity.
\textbf{e)} On the left (right) average Phase Exchange (Fluidity) values for dipeptides categorized according to the hydrophobic (H), polar (P), and charged (C) classes of their two residues. 
Each dipeptide was further separated according to the four RNA bases (A, C, G, U), with each base represented by a distinct color bar.}
\label{fig4}
\end{figure*}

\section{Conclusions}

Our systematic investigation of all 400 dipeptide combinations, both in isolation and in the presence of individual nucleobases, provides a chemically resolved framework for understanding minimal determinants of phase separation. 
We confirm that aromatic and hydrophobic residues preferentially drive the formation of compact, low-exchange peptide clusters, in agreement with previous coarse-grained simulation studies \cite{tang2021prediction,lesniewski2024coarse}. 
Polar and charged residues, in contrast, generally promote higher turnover or prevent phase separation, although specific dipeptide combinations such as LL, VI, EK, LA, and LI display phase separation tendencies consistent with experimental observations \cite{scarel2024self}, highlighting subtle sequence-dependent effects that may be overlooked in coarser models.\\
We further demonstrate that nitrogenous bases modulate dipeptide condensates in a base-specific manner. High-density peptide clusters are often preserved upon nucleobase addition, with their structural organization maintained but their dynamics and lifetime altered. However, for some dipeptides even dense $PCs$ with high persistency can be disrupted by nitrogenous bases, especially when charged residues are considered, reflecting electrostatic interactions with nucleobases. 
Nucleobases can also promote cluster formation, but the resulting $PCs$ tend to be unstable and highly dynamic. These observations parallel cellular mechanisms by which RNA regulates condensate formation, acting both as a nucleation scaffold  \cite{kaur2021sequence,alshareedah2019interplay,alshareedah2020phase} and a modulator of material properties such as fluidity, viscosity, and surface tension \cite{kaur2021sequence,alshareedah2019interplay,alshareedah2020phase,alshareedah2021quantifying,zhang2015rna,wei2017phase}.
Importantly, we find that these effects are strongly dependent on nucleobase identity: adenine tends to produce clusters with higher internal mobility, including for hydrophobic–hydrophobic pairs, while guanine enhances phase exchange in H–H clusters. Pyrimidines (cytosine and uracil) exert comparatively milder and less specific effects.
Overall, our results reveal that even ultrashort peptides encode the essential physicochemical principles of phase separation, and that nucleobase identity provides an additional layer of control over condensate stability and dynamics. This minimal model offers a mechanistic perspective on RNA-mediated regulation of biomolecular condensates and may inform future strategies for tuning or disrupting phase-separated assemblies in therapeutic contexts.

\section{Materials and Methods}

\subsection*{Molecular dynamics simulations}
Initial configurations were generated starting from a single dipeptide structure. Atomic coordinates were aligned by principal component analysis, and the main molecular axis was oriented along the laboratory z-direction.
Replication was performed by translating the aligned dipeptide coordinates on a regular lattice, yielding a homogeneous initial distribution of 360 dipeptides within the simulation box.
All simulations were performed using Gromacs~\cite{gromacs}.
Topologies of the system were built using the CHARMM-27 force field~\cite{mackerell1998all}.
Each system was placed in a 6$\times$6$\times$18.5 $nm$ simulative box, with periodic boundary conditions, filled with TIP3P water molecules~\cite{Jorgensen1983}. 
To study the effect of nitrogenous bases on the dipeptides phase separation propensity, in each box 36 water molecules were substituted with purines (A or G) or pyrimidines (C or U), to obtain a peptide-base ration of 1-10.
For all simulated systems, we ensured that each atom of the proteins was at least 1.1 $nm$ from the box borders.
Each system was then minimized using the steepest descent algorithm. Next, a relaxation of water molecules and thermalization of the system were performed in NVT and NPT ensembles, each for 0.1 $ns$ at 2 $fs$ time-step. 
To relax the initial crystal-like structure, we performed for each dipeptide system a simulation at temperature $T=$380 $K$ of $\sim$50 $ns$. The resulting system was then simulated for additional $\sim$50 $ns$ at $T=$300 $K$. This procedure was followed for all the simulations.

\subsection*{Principal Component Analysis}
Principal Component Analysis (PCA) is a multivariate statistical technique used to reduce the dimensionality of a dataset. 
This is achieved by transforming the original set of correlated variables into an orthogonal basis defined by the eigenvectors of the covariance matrix $\hat{C}$ computed from the observables. These eigenvectors, referred to as principal components (PCs), are ordered according to their corresponding eigenvalues, which quantify the amount of variance captured by each component.
By retaining only the first $d$ PCs it is possible to reduce the dimensionality of the dataset while preserving most of its relevant information.\\
A quantitative measure of the contribution of each PC, associated with eigenvalue $\lambda_i$, is provided by the Explained Variance Ratio (EVR):

\begin{equation}\label{e evr}
    EVR(\lambda_i)=\frac{\lambda_i}{\sum_j^{N} \lambda_j},
\end{equation}
where $N$ is the number of original observables.\\

The contribution of each original variable to a given principal component is described by the loadings, defined as the components of the corresponding eigenvector. Variables with larger absolute loadings contribute more significantly to the variance captured by the corresponding PC.

\subsection*{Structure factor}
The mesoscale organization of the systems were quantified through the static structure factor $S(q)$, defined as
\begin{equation}
    S(\vec{q}) = \frac{1}{N}\langle|\sum_{j=1}^Ne^{i\vec{q}\cdot\vec{r}}|^2\rangle,
\end{equation}
where $N$ is the number of dipeptides in the system and $\vec{r_j}$ is the position of the dipeptide $j$ center of mass.\\
The wavevectors $\vec{q}$ were constructed consistently with periodic boundary conditions as $\vec{q}=2\pi(\frac{n_x}{L_x},\frac{n_y}{L_y},\frac{n_z}{L_z})$, where $L_x, L_y, L_z$ are the box dimensions and $n_x, n_y, n_z$ are integers.
To obtain the isotropic structure factor $S(q)$, wavevectors were grouped into spherical shells of width $\Delta q$. For each shell, $S(q)$ was computed for all wavevectors within the bin and averaged, yielding a radially averaged structure factor.

\subsection*{Morphology characterization of Persistent Clusters}
Clusters composition and extension were compared through the Radius of gyration $R_g$, computed for each cluster at every simulation frame as
\begin{equation}
    R_g = \sqrt{\frac{1}{N}\sum_{i=1}^N|r_i-r_{CM}|^2},
\end{equation}
where $N$ is the number of dipeptides in the cluster, $r_i$
are the dipeptides center-of-mass coordinates, and $r_{CM}$is the cluster center of mass.\\
To demonstrate that $R_g$ is a meaningful descriptor of cluster morphology and spatial organization, we computed the asphericity $a$ from the eigenvalues of the gyration tensor:
\begin{equation}
    a=\frac{3}{2}\frac{\sum_{i=1}^3(\lambda_i-\lambda)^2}{(\sum_{i=1}^3\lambda_i)^2},
\end{equation}
where $\lambda_i$ are the eigenvectors of the covariance matrix and $\lambda=\frac{\sum_{i=1}^3\lambda_i}{3}$.
In most cases, the asphericity remains below 0.3 (see Supplementary), indicating that clusters are predominantly quasi-spherical.

\section*{Author contributions statement}
M.M. conceived the research.  M.M. and G.R  supervised the research. G.G. performed research. All authors analyzed results, wrote, and revised the paper.

 .

\section*{Acknowledgements}
This research was partially funded by grants from ERC-2019-Synergy Grant (ASTRA, n. 855923); EIC-2022-PathfinderOpen (ivBM-4PAP, n. 101098989); Project `National Center for Gene Therapy and Drugs based on RNA Technology' (CN00000041) financed by NextGeneration EU PNRR MUR—M4C2—Action 1.4—Call `Potenziamento strutture di ricerca e creazione di campioni nazionali di R\&S' (CUP J33C22001130001).
Project FIS2023-02957 (CUP B53C24009530001), 'Fathoming oUt the role of partitioninG noIse in cancer epithelial-mesenchymal Transitions (FUGIT),  funded with the contribution of the Italian Ministry of University and Research pursuant to Ministerial Decree No. 1236 of August 1, 2023 - FIS 2 CALL.

\bibliographystyle{unsrt}
\bibliography{mybibfile}

\end{document}


\section{Supplementary Information}

\subsection{Persistent cluster morphology description}
\begin{figure*}[]
\centering
\includegraphics[width=\linewidth]{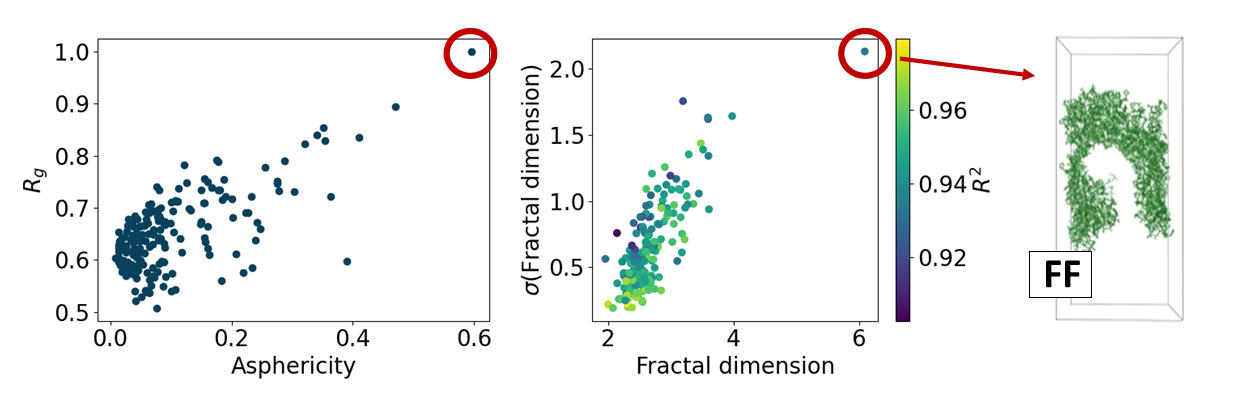}
\caption{\textbf{Persistent Clusters (PCs) as spheres.} 
For each $PC$, Radius of gyration ($R_g$}) versus Asphericity on the left and 
Fractal dimension versus its standard deviation on the right. In the right panel, points are displayed using a color gradient reflecting the associated $R^2$ values. 
The FF system (red circle) is distinguished by its markedly anisotropic, crescent-like morphology.
A representative final configuration of its evolution is shown on the right.
\label{sup1}
\end{figure*}

We computed the Asphericity for all Persistent Clusters ($PCs$) and found it remained below 0.3 for approximately 95\% of them (see Figure~\ref{sup1}), indicating that the aggregates are predominantly quasi-spherical. The value obtained for the FF system ($\sim 0.6$) is reported for completeness; however, its interpretation is not physically meaningful due to the highly anisotropic, c-shaped morphology of the corresponding $PC$, as discussed in the Main Text.
A positive correlation between $R_g$ and Asphericity (Pearson $r = 0.68$, $p < 10^{-26}$) is observed, suggesting that increases in cluster extension are accompanied by moderate shape distortions rather than extreme anisotropic deformations. This supports the use of $R_g$ as a reliable descriptor of aggregate size and morphology within the quasi-spherical regime.\\

\begin{figure*}[]
\centering
\includegraphics[width=\linewidth]{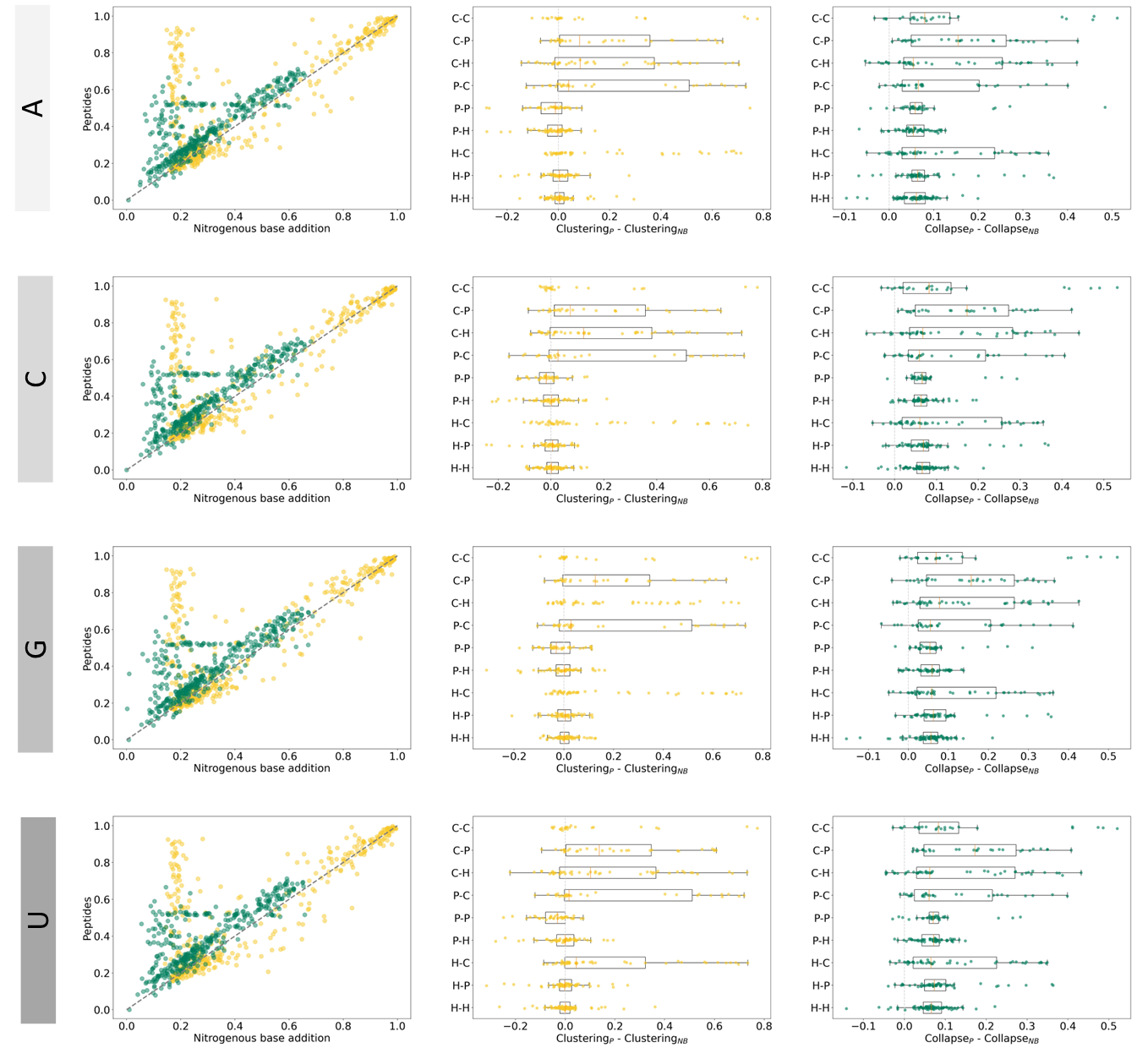}
\caption{\textbf{Nucleobase-specific and residue class-specific modulation of dipeptides clustering.} 
In the first column, comparison of Clustering Degree and Collapse Degree in the presence (x-axis) and absence (y-axis) of arginine (A), citosine (C), guanine (G), and uracil (U), from top to bottom.
The dashed line indicates equality. 
In the second (third) column distribution of changes in Clustering Degree (Collapse Degree) upon nitrogenous bases addition, stratified by residue classes ($C$ = charged, $P$ = polar, $H$ = hydrophobic). Positive values indicate enhanced association in the presence of nucleobases.}
\label{sup2}
\end{figure*}

Having established the validity of the quasi-spherical approximation, we next evaluated the internal structural compactness of $PCs$ by computing their Fractal dimension. Figure~\ref{sup2} reports the Fractal dimension averaged over all frames for each $PC$, together with its standard deviation $\sigma$. The values of Fractal dimension display substantial temporal fluctuations, reflected in relatively large $\sigma$ for most $PCs$.
To assess whether these fluctuations arise from poor mass–size scaling fits, we computed the coefficient of determination $R^2$, a standard measure of goodness-of-fit defined as
\begin{equation}
    R^2=1-\frac{SS_{res}}{SS_{tot}},
\end{equation}
where $SS_{res}=\sum_{i=1}^N(y_i-y_i^{fit})^2$ is the residual error and $SS_{tot}=\sum_{i=1}^N(y_i-\frac{\sum_{i=1}^Ny_i}{N})^2$ is the total data variance.
The lowest observed $R^2$ value is 0.90, while approximately 90\% of $PCs$ exhibit $R^2 > 0.93$, confirming the robustness of the mass–size scaling procedure. The pronounced fluctuations in the fractal dimension are therefore likely attributable to genuine frame-by-frame rearrangements in the internal spatial distribution of $PC$ components, rather than fitting artifacts.

\subsection{Nucleobase-specific modulation of dipeptides phase separation propensity and clusters material properties}

\begin{figure*}[]
\centering
\includegraphics[width=\linewidth]{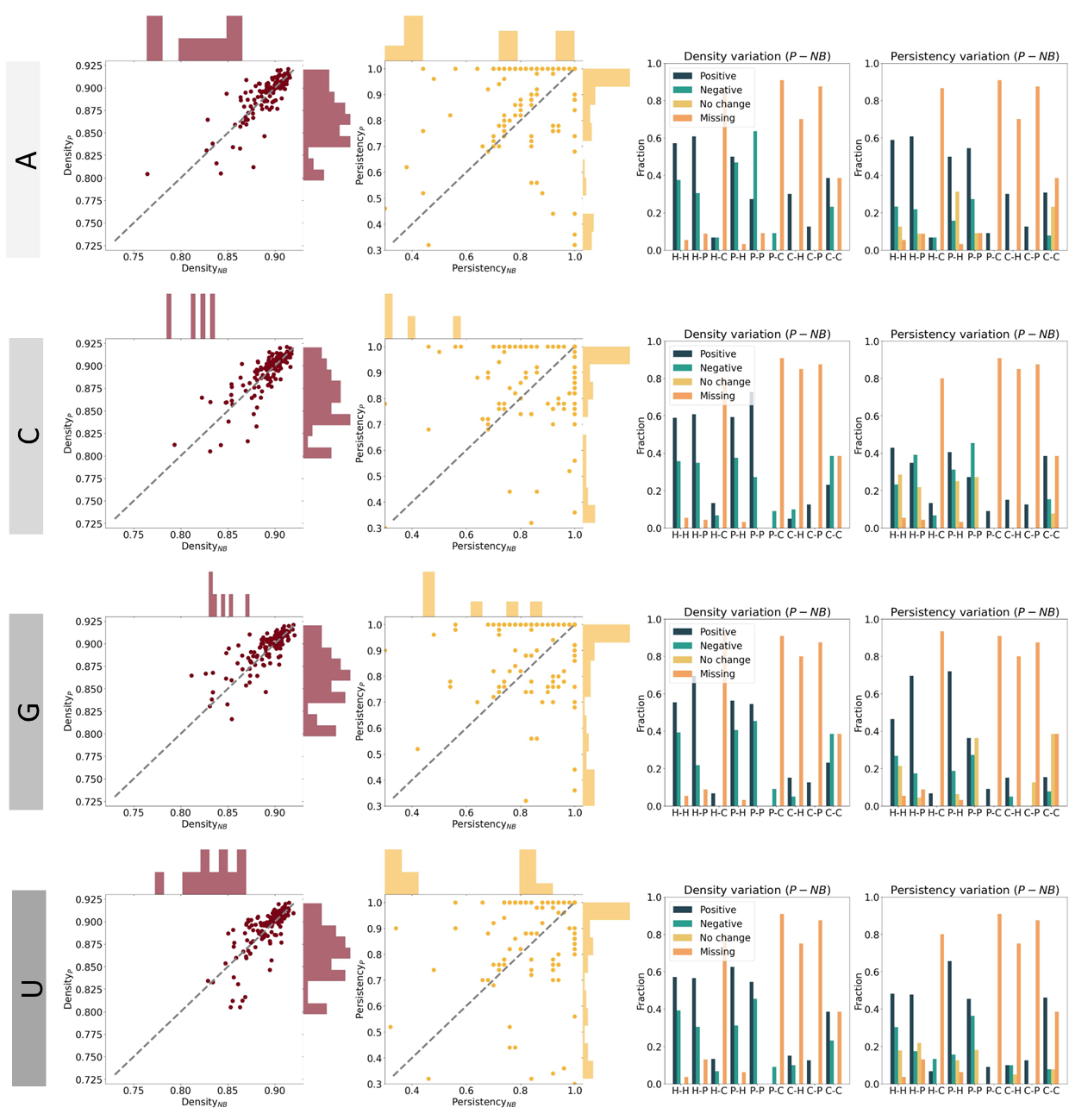}
\caption{\textbf{Nucleobase-specific and residue class-specific modulation of condensates density and stability.}
In the first (second) column, comparison of Density (Persistency) in the presence (x-axis) and absence (y-axis) of arginine (A), citosine (C), guanine (G), and uracil (U), from top to bottom.
The dashed line indicates equality.
The marginal distribution on the top (right) axis indicate the values of the $PCs$ forming only in the systems with (without) nitrogenous base.
In the third (fourth) column categorical variation of Density (Persistency) across residue classes in the presence of each nitrogenous base.
Dipeptides system that completely lose the propensity to phase separate following nitrogenous bases addition are indicated by the orange bar ("Missing").}
\end{figure*}

\begin{figure*}[]
\centering
\includegraphics[width=\linewidth]{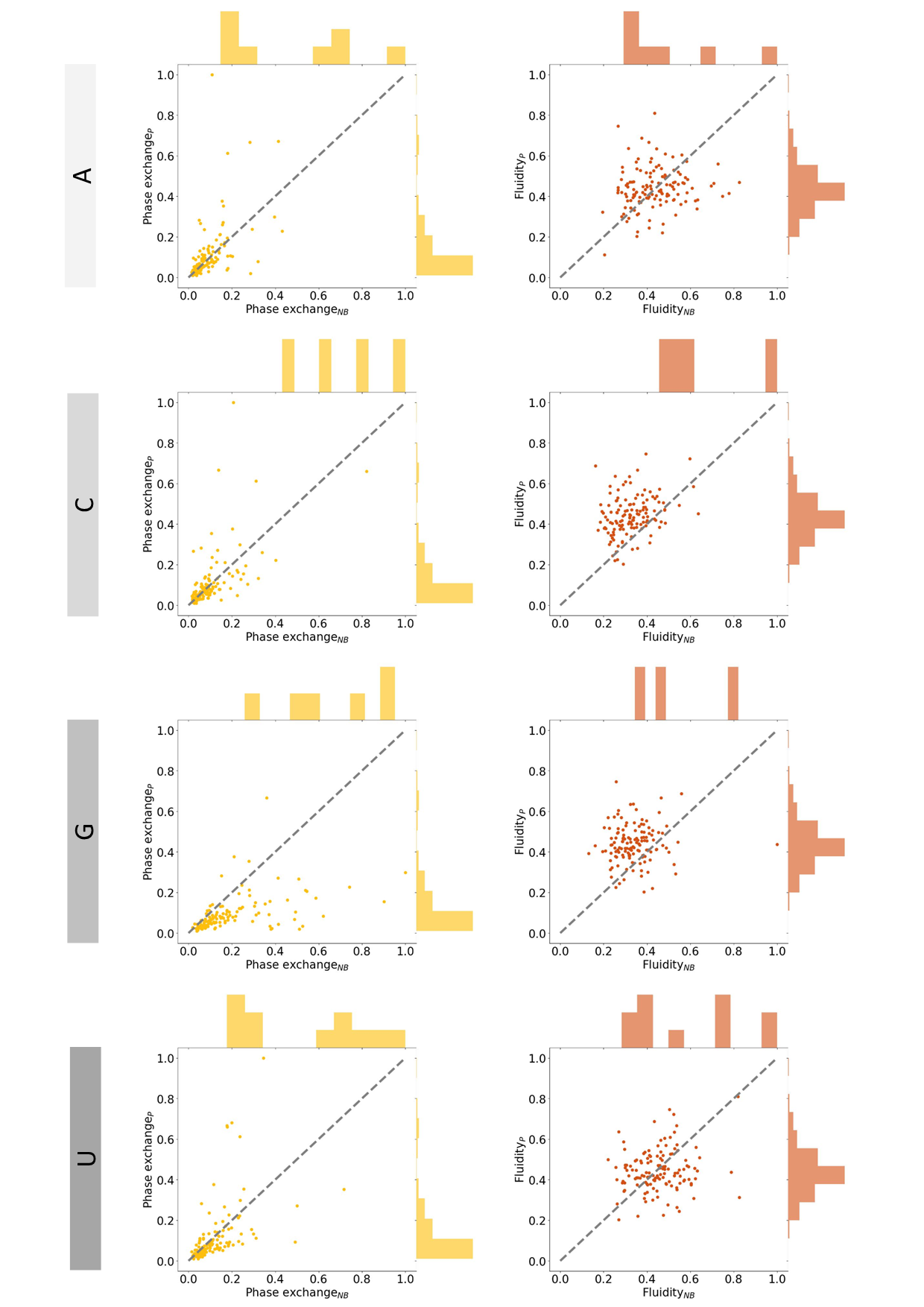}
\caption{\textbf{Nucleobase-specific modulation of condensates exchange and internal dynamic.}
In the first (second) column, comparison of Phase exchange (Fluidity) in the presence (x-axis) and absence (y-axis) of arginine (A), citosine (C), guanine (G), and uracil (U), from top to bottom.
The dashed line indicates equality.
The marginal distribution on the top (right) axis indicate the values of the $PCs$ forming only in the systems with (without) nitrogenous base.}
\end{figure*}